\documentclass[linenumbers]{aastex631}

\usepackage{gensymb}
\usepackage{amsmath}

\begin{document}
\nolinenumbers
\title{Proposed Resolution to the Solar Open Magnetic Flux Problem}

\author[0000-0001-9326-3448]{C. Nick Arge}
\affiliation{NASA Goddard Space Flight Center, 8800 Greenbelt Rd., Greenbelt, MD 20771, USA}

\author[0000-0002-4752-7534]{Andrew Leisner}
\affiliation{George Mason University, Fairfax, VA, USA}

\author[0000-0002-1091-4688]{Samantha Wallace}
\affiliation{NASA Goddard Space Flight Center, 8800 Greenbelt Rd., Greenbelt, MD 20771, USA}

\author[0000-0002-6038-6369]{Carl J. Henney}
\affiliation{AFRL/Space Vehicles Directorate, Kirtland AFB, NM, USA}



\begin{abstract}
The solar magnetic fields emerging from the photosphere into the chromosphere and corona are comprised of a combination of “closed” and “open” fields. The closed magnetic field lines are defined as those having both ends rooted in the solar surface, while the “open” field lines are those having one end extending out into interplanetary space and the other rooted at the Sun’s surface. Since the early 2000’s, the amount of total unsigned open magnetic flux estimated by coronal models have been in significant disagreement with in situ spacecraft observations, especially during solar maximum. Estimates of total open unsigned magnetic flux using coronal hole observations (e.g., using extreme ultraviolet (EUV) or Helium (He) {\rm I}) are in general agreement with the coronal model results and thus show similar disagreements with in situ observations. While several possible sources producing these discrepancies have been postulated over the years, there is still no clear resolution to the problem. This paper provides a brief overview of the problem and summarizes some proposed explanations for the discrepancies. In addition, two different ways of estimating the total unsigned open magnetic flux are presented, utilizing the Wang-Sheeley-Arge (WSA) model, and one of the methods produce surprisingly good agreement with in situ observations. The findings presented here suggest that active regions residing near the boundaries of mid-latitude coronal holes are the probable source of the missing open flux. This explanation also brings in line many of the seemly contradictory facts that have made resolving this problem so difficult.

\end{abstract}



\section{Introduction} \label{Sec_1:Intro}
Wang and Sheeley \citep{wang1995, wang2000} were the first to systematically compare the total open unsigned solar magnetic flux as determined from a potential field source surface (PFSS) model \citep{schatten1969, altschuler1969, wang1992} with that derived using in situ spacecraft observations. The comparison, which was actually between average model field strengths and in situ observations at 1 au, spanned a 28-year time interval and showed good overall agreement between the two quantities. While there were periods of noticeable disagreement (e.g., 1985\,--\,1989 and 1997\,--\,1999), difficulties with measuring the polar fields were considered a possible source of the problem. However, a subsequent paper by \citet{wang2002} showed the model and observations beginning to diverge from one another starting around 1997. By early 2005 Yi-Ming Wang had personally conveyed to the lead author of this paper that the disagreement persisted and that it was not clear what the problem was. Two decades later, model and in situ derived unsigned open magnetic flux values have yet to come back into alignment and show discrepancies as large as factors of two or more. The heliospheric community has been attempting to resolve the problem ever since. 

Recent literature is rich in detailed overviews of what is now referred to as the “open flux problem” and the reader is especially encouraged to review papers by \citet{wang2022}, \citet{badman2021}, \citet{wallace2019}, \cite{linker2017,linker2021}, and \cite{lowder2017} as well as references therein. Here we provide a brief summary of the attempts to understand and resolve the problem. In general, it comes down to whether the issue lies with the models, the input data sources to the models (i.e., global photospheric magnetic field maps), the in situ observations, or some combination of the three. Since the sustained discrepancy between the model and in situ derived unsigned open magnetic fluxes started just as solar activity began to pick up after the 1996 minimum, \citet{riley2007} and \cite{owens2006} independently suggested that the missing flux might be from coronal mass ejections (CMEs) that had yet to disconnect from the Sun. It is not straightforward to distinguish closed magnetic field lines from open ones in in situ observations when estimating total unsigned heliospheric magnetic flux, and PFSS-based coronal models cannot account for closed magnetic flux from CMEs that have yet to disconnect from the Sun. While this explanation appeared to be a plausible (though highly speculative) solution, it has largely fallen out of favor. For example, \citet{wang2022} suggested this is not the major source of the problem, though possibly a contributor. Other problems became apparent over time with the in situ data such as the discovery that heliospheric flux derived from them increased as function of distance from the Sun \citep{owens2008}. However, they showed this was not a major issue until beyond the orbit or Mars. \citet{badman2021} reviews the history of efforts exploring problems with estimating the total heliospheric magnetic flux from single point in situ observations (e.g., velocity shears that warp the field and local field inversions) and methods for more accurately estimating it. They also show, using Parker Solar Probe (\textit{PSP}: \citealp{Fox2016}) data, that $\mathbf{|B_{r}|R^{2}}$ remains constant for measurements spanning radial distances from the Sun between 0.13\,--\,0.8au, latitudes within 4{\degree} of the ecliptic plane, and longitude, which is in alignment with that originally found by \citet{smith1995, Balogh2001, smith2003}. This critical finding by Smith and Bologh, discussed in greater depth below, allows total heliospheric magnetic flux to be determined using single point in situ measurements. \citet{badman2021} concludes that total unsigned open magnetic flux from the Sun should be equivalent to the total heliospheric unsigned magnetic flux as derived by in situ observations (at least starting from about 0.13au from the Sun) and serves as key constraint to coronal models. 

While routine measurements of the Sun’s photospheric magnetic field have been made for more than 50 years, the measurements remain highly challenging to calibrate and subject to potentially large uncertainties. Yet, essentially all coronal models require global maps of the photospheric magnetic field constructed from full-disk magnetograms as their input. Problems abound with the global maps assembled from these observations such as lack of observations on the solar far side, unreliable measurements near the observed solar limbs \citep[e.g.,][]{Harvey2007} and thus at the poles, calibration of the measurements (both absolute as well as intercalibration between different magnetographs), and assumptions made to obtain the radial photospheric magnetic field component. \citet{posner2021} describe several of the key problems with the photospheric magnetic field maps used to drive coronal models and how they can impact solutions of the global coronal field. Global photospheric magnetic field maps have therefore been considered a prime suspect producing the disagreement between model and in situ derived total unsigned open magnetic flux. \citet{wang2022}, \citet{wallace2019}, \citet{riley2014}, and \citet{arge2002} describe in detail several specific problems with global photospheric maps and how they can impact model estimates of the total open flux. In short, it is not a simple matter of the magnetic field maps needing to be multiplied by a fixed offset, as this would not explain, for example, why in situ and model open flux values agree so well until the late 1990s. There has been speculation that the source of the missing flux resides near the Sun’s poles \citep[e.g.,][]{linker2017}, where the field values are very uncertain and model solutions highly sensitive to their strength \citep{wang2022,posner2021}. 

Besides observational input data challenges, coronal models themselves are a possible source of the open flux problem and could lack the required physics necessary to reproduce the proper amount of open magnetic flux or key model parameters may be poorly determined or even vary. For example, \citet{badman2021}, \citet{Arden2014}, \citet{Meadors2020}, and \citet{lee2011} explored varying the source surface height in PFSS models to improve agreement with the observed open flux based on in situ observations. As pointed out by \citet{wang2022}, this solution is problematic, as varying the source surface height also changes the area of coronal holes. Increasing the amount of open flux requires lowering the source surface height, which results in increased coronal hole size. However, \citet{wallace2019} showed (on three Carrington rotation time scales) that the total unsigned open flux as determined by the potential field based Wang-Sheeley-Arge (WSA) model with a fixed source surface height of about 2.5$R_{s}$ agrees well, except near solar maximum, with open magnetic flux independently determined using observationally-derived coronal hole areas and the identical photospheric magnetic field maps used to drive the WSA model. Lowering the source surface height would therefore likely generate coronal holes far too large. In general, models used to calculate open flux are either magnetostatic PFSS models or magnetohydrodynamic (MHD) models run to steady state \citep[e.g.,][]{detoma2005}. Both approaches produce coronal solutions where every point on the solar surface is unambiguously open or closed. Yet, the Sun’s surface and corona magnetic fields evolve over time and therefore are highly dynamic in nature, with certain regions (e.g., coronal hole boundaries and active regions) having the magnetic field repeatedly switch from open to closed via, for example, interchange reconnection. PFSS models are incapable of modeling this behavior of the Sun, while doing this with more advanced coronal models has proven extremely challenging due to the difficulties with the properly updating the photospheric field boundary maps used to drive them in a time-dependent manner. This time-dependent behavior could be a source of the missing open flux. 

This paper is a follow on to \citet{wallace2019}. The same coronal model (WSA) and photospheric field input Carrington maps from the National Solar Observatory (NSO) are used to generate the coronal solutions and therefore model derived total open unsigned magnetic fluxes. However, the coronal open flux estimates have been extended to 2017 when the Synoptic Optical Long-term Investigations of the Sun (SOLIS)/Vector Spectromagnetograph (VSM) instrument \citep{henney2006,henney2009} magnetograph recorded its last magnetograms before being shut down to be relocated. The same open unsigned magnetic heliospheric flux values based on in situ spacecraft measurements are used here as well (see \citealt{wallace2019} for details). Here we attempt to reconcile several of the seeming contradictory findings and results that have emerged over the last 20 years and present new results that appear to explain the source of the missing open insigned solar magnetic flux. In Section \ref{Sec_2:WSA_model}, we describe the WSA model. Section \ref{Sec_3:Methods} describes two methods for calculating total open unsigned magnetic flux from the WSA model. Section \ref{Sec_4:Discussion} provides a detailed discussion of the results and their implications. A summary of the results is provided in Section \ref{Sec_5:Summary}. 

\section{The WSA Model} \label{Sec_2:WSA_model}
The coronal model used in this study is the Wang-Sheeley-Arge (WSA) model \citep{arge2000, arge2003, arge2004, mcgregor2008, wallace2019}. The coronal portion of WSA is comprised of two potential field (PF) type models. The inner model is a traditional PFSS model, it specifies the coronal field from the inner, photospheric boundary at 1$R_{s}$ to its outer boundary or source surface. In this study, the source surface is set to 2.51$R_{s}$, which is also the radius used in \citet{wallace2019}. The second PF model used is the Schatten Current Sheet (SCS) model \citep{schatten1971}. It provides a more realistic description of the outer coronal field. A small overlap region between the PFSS and SCS solutions is used to interface the two models in WSA, where the radial magnetic field components of the PFSS magnetic field solution at 2.49$R_{s}$ are used as the inner boundary condition to the SCS model. In this implementation, all magnetic fields beyond 2.49$R_{s}$ are, by definition, open (see \citealt{mcgregor2008}). Formally, the outer boundary of the SCS model extends out to infinity, but only the portion of the model out to between 5$R_{s}$ and 21.5$R_{s}$ is typically used and can be set by the user. This outer surface is referred to as the outer boundary of the model. The radius chosen depends on how the model is used. A value 21.5$R_{s}$ is often used when the magnetic field output of the SCS is used to drive advanced MHD solar wind models such as Enlil \citep{odstrcil2004,odstrcil2005} and Gamera \citep{zhang2019, merkin2016}. This radius virtually guarantees that the solar wind is both everywhere supersonic and super alfvenic, which makes the inner boundary conditions for MHD solar wind models much simpler to manage. For the 1D kinematic solar wind model often used in WSA \citep{arge2004}, 5$R_{s}$ is commonly used. While the outer boundary is set to 5$R_{s}$ in this study, any value can be used, as the total unsigned open flux remains constant on any surface beyond 2.49$R_{s}$. 

\section{Methods for calculating total open flux with the WSA model} 
\label{Sec_3:Methods}

\begin{figure}[ht!]
\plotone{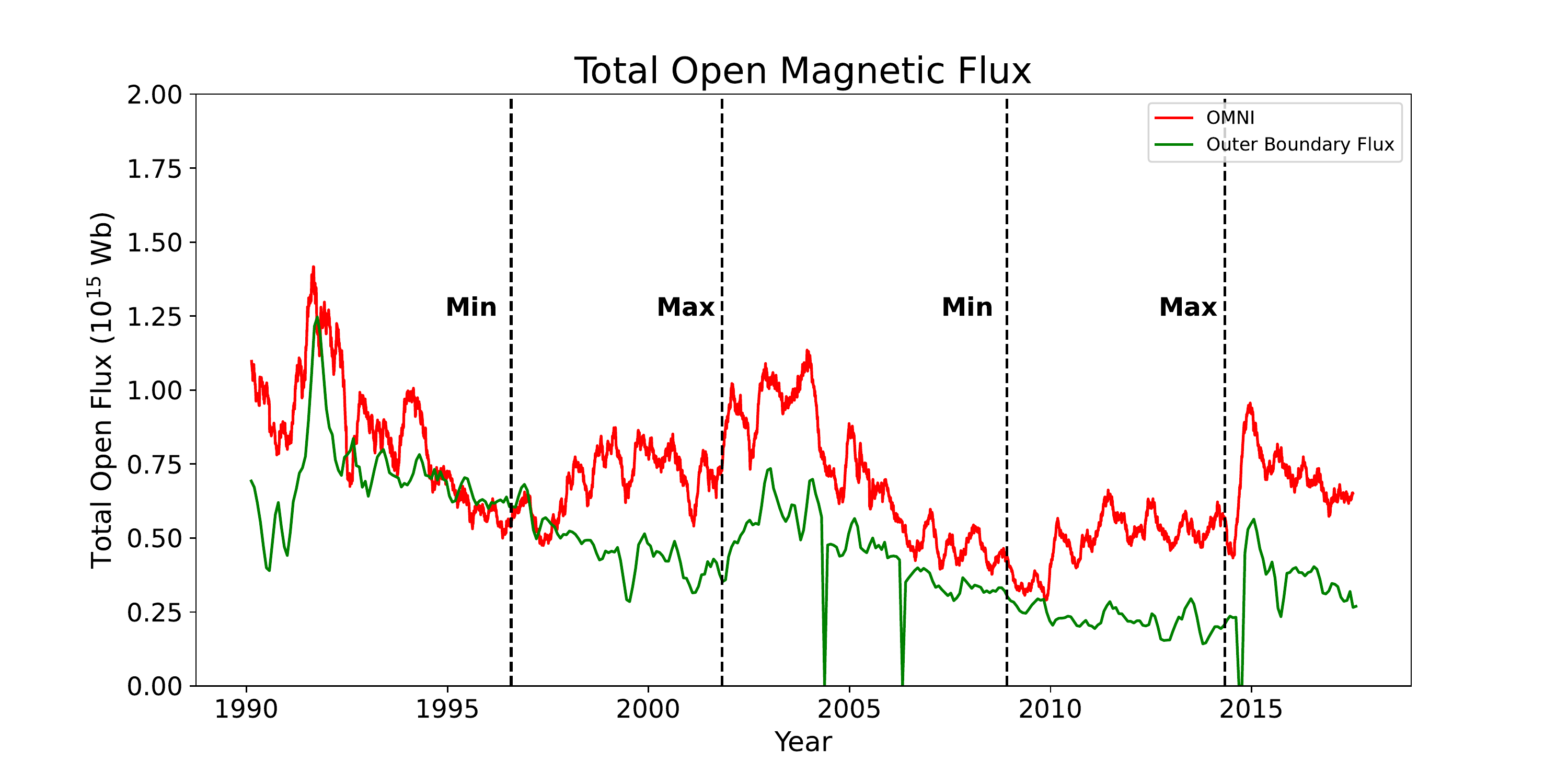}
\caption{Total open magnetic flux based on summing the unsigned magnetic flux in each grid cell on the outer boundary (R=5.0$R_{s}$) of the WSA model (green) and using in situ (OMNI) observations (red). The values shown are for 3 Carrington rotation running averages. Values of zero in the model results occur because of data gaps in photospheric magnetic field maps used as input to the WSA model. The vertical dotted lines indicate the times of solar minimum and maximum. 
\label{Fig.1}}
\end{figure}

\subsection{Traditional Approach} \label{Sec_3.1:Trad_Appr}
The traditional way of calculating the total open unsigned magnetic flux using a PFSS model is to sum the unsigned magnetic flux in each grid cell on the source surface of the model. As mentioned above, WSA uses both the PFSS and SCS models, so the total unsigned open magnetic flux is calculated on the outer boundary of the model located at 5$R_{s}$ in this work. The results from this approach are shown in Figure~\ref{Fig.1}, where the values plotted are the running averages using a temporal window of three Carrington rotations ($\sim$82 days).

Figure~\ref{Fig.1} also shows the three Carrington rotation averaged total open unsigned magnetic flux as determined using in situ spacecraft measurements. The in situ data shown here is the same data set as described in \citet{wallace2019}, except that it has been extended to 2017. Note that from this point forward the expression “open flux” will be used as shorthand for total open unsigned magnetic flux unless clearly indicated otherwise. Using Ulysses observations, \citet{smith2003} demonstrated $\mathbf{|B_{r}|R^{2}}$ was a constant during two of the fast latitude scans (i.e., one during solar minimum and another at maximum), where $\mathbf{R}$ is the radial distance from the Sun to the spacecraft, and $\mathbf{|B_{r}|}$ is the absolute value of the measured radial magnetic field at the spacecraft. By multiplying this result by 4$\pi$, one obtains the global heliospheric flux. \citet{lockwood2009} and \citet{owens2017} demonstrated that as long as daily average values are used, this approach works reasonably well. \citet{owens2008} demonstrated for multiple spacecraft positioned inside Mars' orbit that they all provided very similar results for the total unsigned open flux. Of course, this just means that they give self-consistent results and not necessarily the correct value. As can be seen in Figure~\ref{Fig.1}, there are clear discrepancies between the unsigned open fluxes values as determined by in situ methods and that provided by summing the open magnetic flux on the outer boundary of the WSA model. The differences, which can exceed factors of two, are persistent except near periods of solar minimum (e.g., during the years $\sim$1996 and $\sim$2009). As explained in the introduction, the source(s) producing these discrepancies has(have) been a longstanding mystery.

\subsection{Using Field Line Tracing} \label{Sec_3.2:FL_tracing}

\begin{figure}[p]
\plotone{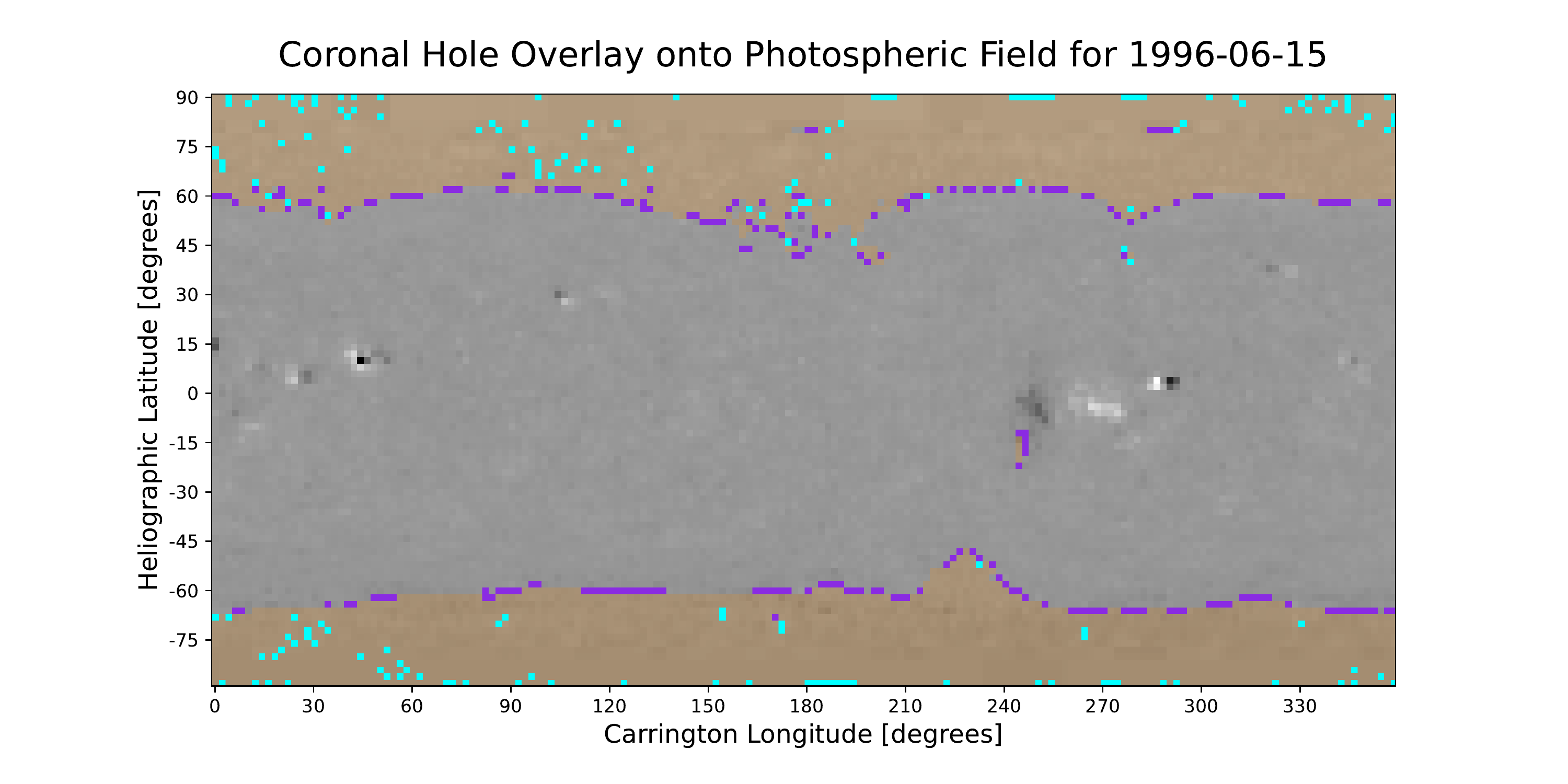}
\plotone{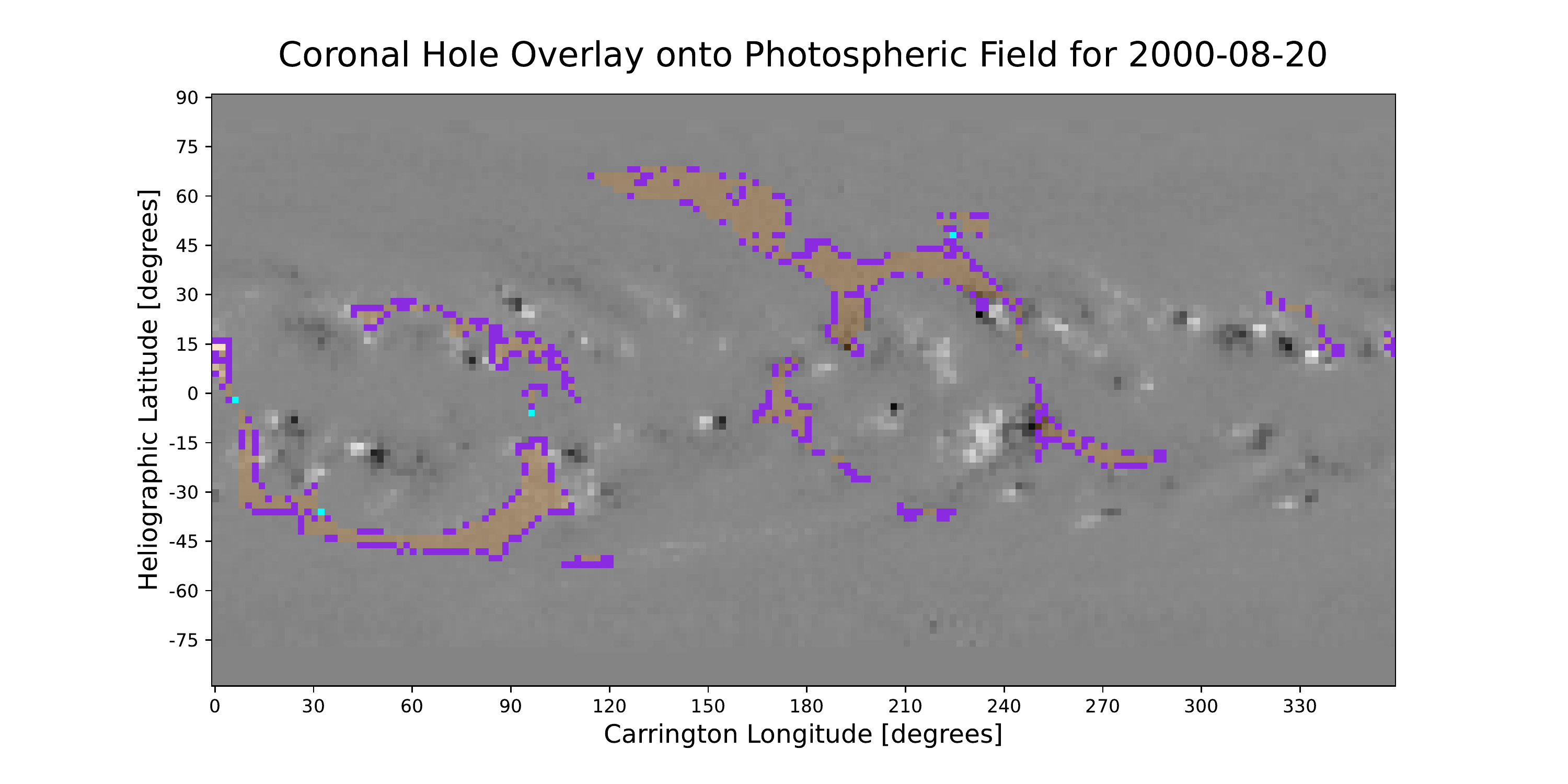}
\caption{(Top) SOLIS/VSM photospheric magnetic field map for Carrington Rotation (CR) 1910 with the gray scale indicating magnitude, and white/black delineating the positive/negative magnetic field polarity. CR1910 occurred near solar minimum. Orange shading indicates bi-directional open cells. The outward and inward only open cells are plotted in aqua and purple, respectively. Note that the the inward only cells in purple lie primarily near coronal hole boundaries, while the outward only cells in aqua lie mostly inside coronal holes. (Bottom) Same as above but for CR1966, which occurred during solar maximum. Note how the inward only cells in purple often lie in close proximity to active regions. 
\label{Fig.2}}
\end{figure}

Another approach for calculating total unsigned open flux is to trace magnetic field lines down from the outer boundary of the WSA model to the photosphere to identify magnetically open regions (i.e., coronal holes). This is the approach taken by \citet{lowder2017} and referred to as \textit{inward} tracing in this paper. More specifically, starting at the center of each cell on the outer boundary, the magnetic field is traced down to the photosphere. Cells on the photosphere (i.e., R=1$R_{s}$) that the traced magnetic field lines land in are assigned an open status, while the rest are assigned a closed status. The open cells define the model derived coronal holes. The total open flux is then determined by calculating the unsigned magnetic flux in each of the magnetically open cells and summing the results. Note, the unsigned magnetic flux in each magnetically “open” grid cell is only counted once, even if more than one traced magnetic field line falls within it. 

While tracing magnetic fields from the outer boundary of the model down to the photosphere is the traditional approach for determining magnetically open regions on the Sun, it is well known that this approach occasionally misses open grid cells on the photosphere because of the finite number of magnetic field lines traced (i.e., because of under sampling). A common approach to remedy this is to trace the magnetic field in the opposite direction, that is, from the photosphere out into the corona. This approach is referred to as \textit{outward} tracing in this paper. In particular, starting at the center of each grid cell on the photosphere, the magnetic field is traced outward, and if it reaches the source surface (more precisely  2.49$R_{s}$, due to the way the PFSS and SCS are interfaced in WSA), the field is open by definition, otherwise it is closed. When this is done, one occasionally finds photospheric grids cells that are magnetically open that were missed using inward tracing. 


Figure~\ref{Fig.2} shows two periods (i.e., one during low and the other during high solar activity) in which grid cells on the photosphere were determined to be magnetically open based on both the inward and outward tracing methods. The open cells are overlaid onto a corresponding map of the observed radial photospheric magnetic field (i.e., gray scale portion). Cells determined to be open exclusively using inward tracing appear as purple cells, while those for exclusively outward tracing are aqua. Those identified as open using both methods are shaded in orange and represent the vast majority of the open cells. They are referred to as \textit {bi-directional} open cells in this paper. As can be seen, the exclusively inwardly traced field lines (i.e., purple cells in Figure~\ref{Fig.2}) often lie along the perimeter of the coronal holes and near active regions. We refer to these cells from hereon as \textit{perimeter} cells. One also sees a smattering of exclusively outwardly traced field line (i.e., aqua cells in Figure~\ref{Fig.2}), which can be located anywhere, but most are located at the interiors of coronal holes. 

\begin{figure}[ht!]
\plotone{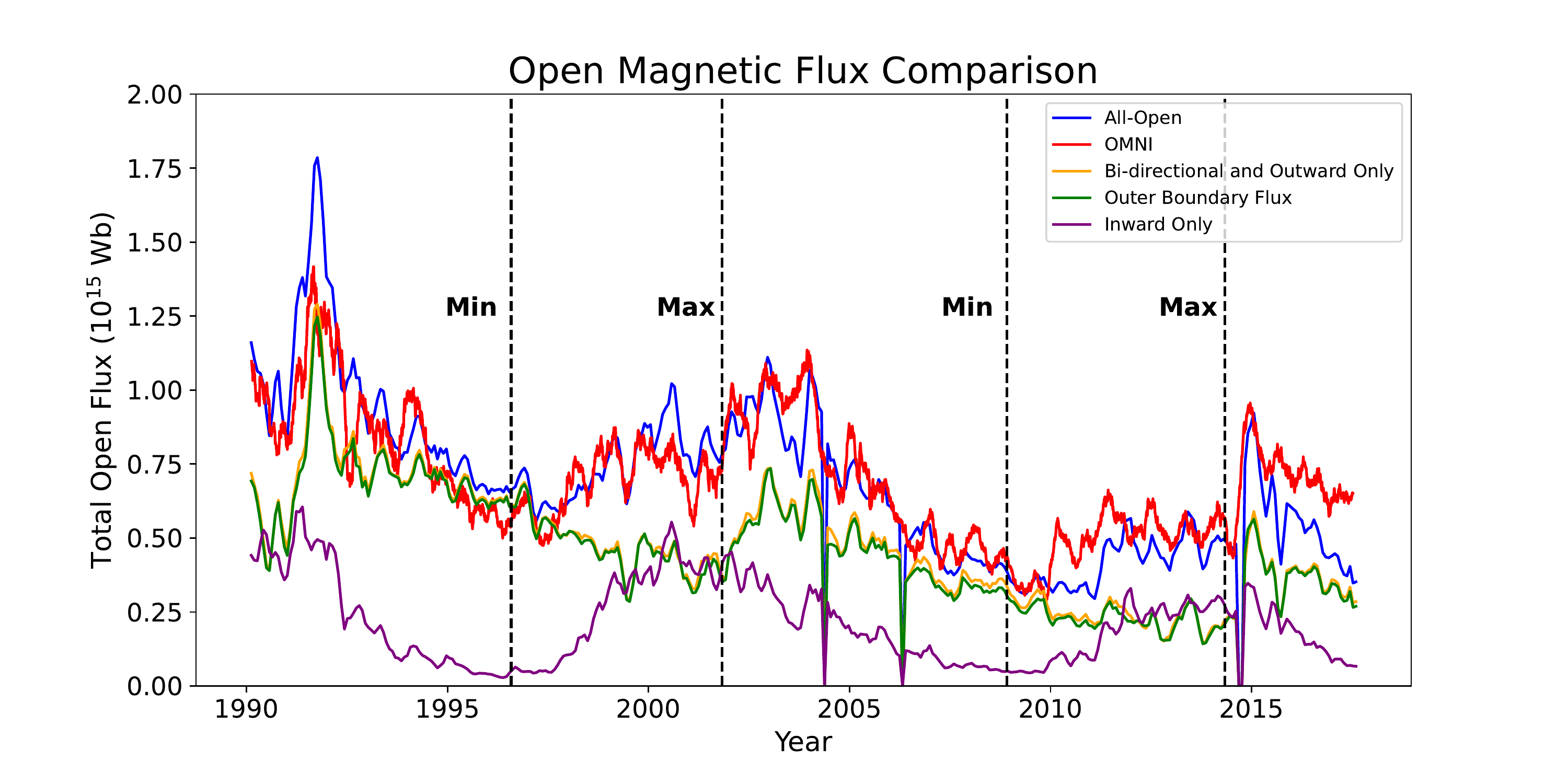}
\caption{Total unsigned open magnetic flux estimated using a variety of approaches: in situ observations (red), bi-directional field line tracing plus outward only tracing (orange line), all-open cells using both inward and outward tracing (blue), inward only tracing (purple), and the traditional approach of calculating the open flux on the outer boundary of the WSA model (green). Note that the inwardly only tracing result (purple) corresponds to the flux derived from only the perimeter cells (purple cells in Figure~\ref{Fig.2}). Further, notice how the total open flux based on summing the bi-directional plus outward only cells (orange) agree well with the traditional method for estimating open flux (green). The all-open cells result (blue) is obtained by summing the flux from bi-directional cells, inward only, and outward only (i.e., by adding the results in orange and purple), and agrees well with the in situ derived open flux (red) suggesting that the missing open flux comes almost exclusively from the fluxes in the perimeter cells. The vertical dotted lines indicate the times of solar minimum and maximum.}
\label{Fig.3}
\end{figure}

Figure~\ref{Fig.3} compares the total open flux for all cells identified as open (from this point on referred to as \textit{all-open} cells) using field line tracing (i.e., for inward, outward, or both) versus that using the open flux derived from the model’s outer boundary and in situ observations. As seen in the figure, the all-open cells magnetic flux estimates agree astonishing well with the in situ results! There are still periods where there are discrepancies (e.g., the intervals between 2010\,--\,2011 and after 2016), but they are minor in comparison to the results obtained when just computing the total unsigned flux on the outer boundary of the model. Has the “missing” open flux been found? 

Also shown in Figure~\ref{Fig.3} are the open magnetic flux results for the combined bi-directional open cells plus those only for outward only tracing (i.e., the fluxes residing in both the aqua and orange color cells shown in Figure~\ref{Fig.2}). The open flux determined from these values agree extremely well with that found using the values on WSA’s outer boundary. Finally, the open flux values for cells identified as open exclusively for inward field line tracing (i.e., the purple cells in Figure~\ref{Fig.2} commonly seen on the perimeter of coronal holes) are shown. It is evident that these cells contribute a significant amount open magnetic flux (upwards of $\sim$50\% near solar maximum) except around the time of solar minimum, which is consistent with the findings of \citet{schrijver2003}. These results are discussed and interpreted in more detail in the following section. 

\section{Discussion}\label{Sec_4:Discussion}
 

The excellent agreement in total unsigned open magnetic flux derived from in situ measurements (Fig.~\ref{Fig.3}, red) and the WSA all-open cells field line tracing method (Fig.~\ref{Fig.3}, blue) is remarkable and almost certainly not due to chance. However, this surprising result raises two key questions. One, what specifically is the source of the missing open solar flux, and two, what is the reason for the disparity between the model open flux estimates based on calculating it on the outer boundary and using magnetic field line tracing? 


\subsection{Source of the Missing Open Heliospheric Flux}
\label{Sec_4.1:Source_of_Missing_OF}

As can be seen in Figure~\ref{Fig.2}, a preponderance of the magnetically open regions identified exclusively by inward tracing reside along the perimeters of coronal holes, and this is independent of solar cycle. During solar maximum, several of these perimeter cells lie very close to strong active regions. The significant contribution of these perimeter cells to the total open flux is shown in Figure~\ref{Fig.3} (purple), accounting for the open magnetic flux needed for the WSA model values to better agree with the in situ measurements. These perimeter cells are also found during solar minimum, but there are far fewer active regions and mid-latitude coronal holes during this time for them to produce significant amounts of magnetic flux. 

\begin{figure}[ht!]
\plotone{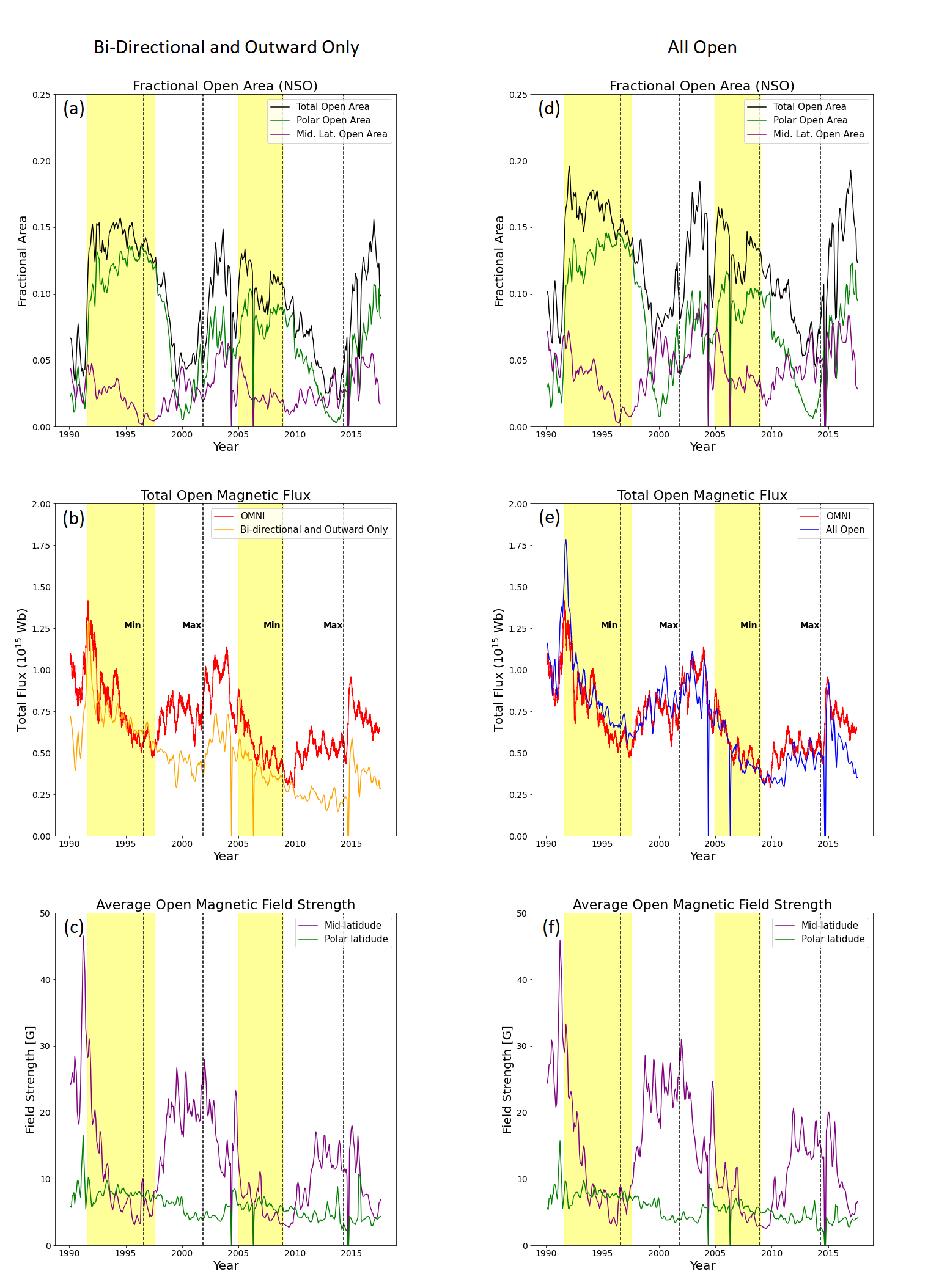}
\centering
\caption{Left panels correspond to results for bi-directional plus outward only cells from 1990 to 2017. Panel 4a corresponds to total fractional open area (black line) and fractional area above (green line) and below (purple line) $|45 \degree |$ latitude. Panel 4b corresponds to the total open flux based on WSA (orange line) and in situ observations (red line). Panel 4c corresponds to the average magnetic field strengths in WSA derived coronal holes above (green line) and below (purple line) $|45 \degree |$ latitude. Results shown in the right panels are identical to those on the left, except they are now for the all-cells results. The vertical dotted lines indicate the times of solar minimum and maximum.}
\label{Fig.4}
\end{figure}

The possibility that magnetically open areas near active regions are the source of the missing open flux is explored further using the results shown in Figure~\ref{Fig.4}. The left column in this figure displays a variety of results for fractional coronal hole area (i.e., total, polar, and equatorial), total open unsigned magnetic flux, and the average field strength in coronal holes for bi-directional plus outward only cells, while the right column shows identical results but for the all-open cells results. More specifically, Figure~\ref{Fig.4}a shows the model determined total fractional coronal area on the Sun and the areas for the mid-latitude regions (i.e., latitudes less than or equal to $|45 \degree |$) and the polar regions (i.e., latitudes greater than $|45 \degree |$). Panel \ref{Fig.4}b dipicts the total open flux derived using the open cells as described just above (note, these fluxes are essentially equal to that found on the outer boundary of WSA, as can been seen in Figure~\ref{Fig.3}{\bf )} and for the in situ measurements, and panel~\ref{Fig.4}c shows the average magnetic field strength residing in the mid-latitude and polar coronal holes. {Panels d-f in Figure~\ref{Fig.4}, show the identical results as panels a-c, but for the all-open cells results. 

As can be seen in Figure~\ref{Fig.4}b, good agreement between the model and in situ determined heliospheric open magnetic fluxes begins in roughly in 1992 where the open flux sharply peaks and then ends in mid-1997. This 5-year period is shaded yellow for clarity.  Figure~\ref{Fig.4}a highlights that the start time of this interval of agreement corresponds to when there is an abrupt increase in the polar coronal hole area and an overall decrease in mid-latitude coronal hole area. The end date begins roughly when the polar holes begin to experience a sustained, and eventually, a substantial drop in area, while the mid-latitude magnetic field strength (Figure~\ref{Fig.4}c) first matches that of the polar regions and then undergoes a subsequent and persistent increase in strength. During the 1992\,--\,1997 period, the average magnetic field strength in the polar coronal holes stays relatively constant, while the strengths in the mid-latitude regions eventually drops below that found in the polar coronal holes. As the polar coronal hole area begins a sustained decline ($\sim$1997) and the magnetic field strength in the mid-latitude holes increases, the discrepancy between in situ and the model derived open heliospheric fluxes is persistent and pronounced. Note that the average polar field strength in Figure~\ref{Fig.4}c and \ref{Fig.4}f stays above zero, which occurs because the average magnetic field strengths are estimated for coronal holes spanning latitudes from $|45 \degree |$ to the poles. The open flux values begin to converge again starting around mid-2005 and then diverge again near the beginning of 2009. During this period, as before, the polar coronal areas are large compared to the mid-latitude values, and the average magnetic field strength in mid-latitude holes are (eventually) generally smaller than in the polar coronal holes. This period (i.e., 2005\,--\,2009) is also highlighted in yellow. While the total open flux values nearly agree during this period, there remains a modest separation in this case. Interestingly, the polar coronal hole fractional area is less pronounced and only about two-thirds that of the previous period when the open flux values showed very good agreement. The mid-latitude coronal hole area also never reaches the very low values (near zero at one point) seen in during the previous period of agreement. The average magnetic field strength in the polar coronal holes is only about 75\% of the values during the first period of good agreement. However, the average field strengths in the equatorial holes are roughly the same for both periods of agreement, at least during the latter part of this interval when the open fluxes match most closely. Clearly, the mid-latitude regions play more of a role, relatively speaking, in the open flux balance during the second period of (near) agreement. Altogether, these results suggest that the significant discrepancies between the in situ and the traditional model method for determining total open unsigned magnetic flux has something to do with the mid-latitude regions on the Sun, especially during higher activity periods. 

As can been seen in Figure~\ref{Fig.4}e, the total unsigned open flux estimated using the all-open cells and in situ observations are in excellent overall agreement. In Figure~\ref{Fig.4}d, the total polar coronal hole area is modestly larger by about 10\,--\,15\% throughout the entire interval shown, compared to that in 4a. However, the equatorial coronal hole areas are significantly larger with increases of approximately 40\% during solar minimum and 80\% during maximum. Yet, caution needs to be applied when interpreting these numbers, as they often correspond to a large increase in what was an originally a small number. For example, just after the year 2000, the fractional mid-latitude coronal hole area increased from $\sim$0.025 in panel \ref{Fig.4}a to $\sim$0.045 in panel 4d. In Panel \ref{Fig.4}f, the average magnetic field strengths in the polar regions are slightly reduced (from $\sim$2\,--\,5\%) throughout the entire time considered for the all-open cells compared to that in \ref{Fig.4}c. The mid-latitude average magnetic field strengths are basically unchanged in Figure~\ref{Fig.4}f compared to \ref{Fig.4}c for the solar minimum periods. However, the values are upwards of 20\% larger in panel~\ref{Fig.4}f for periods outside of solar minimum, especially near solar maximum. The increases in both area and average magnetic field strength in the mid-latitude coronal holes outside solar minimum clearly combine to bring the model derived open fluxes into good agreement with the in situ derived heliospheric open flux values throughout the 27-year period of comparison. All of the evidence described above strongly suggests that the source of the missing open magnetic resides in magnetically open areas that lie near strong field regions. It also explains why attempts to estimate the total open flux using coronal hole boundary estimates derived from EUV and He {\bf{\rm I}} 1083.0 nm data \citep{wallace2019, linker2021} also do not reveal the source of missing open flux, as active regions are bright in EUV and dark in He {\bf{\rm I}} 1083.0 nm and hence open flux near them are not included. 

\subsection{Disagreement Between Open Flux Calculated at the Outer Boundary and Using Magnetic Field Line Tracing}\label{Sec_4.2:Disagreement}

The next question to be addressed is the apparent conundrum of why WSA (or any PF type model) provides different open flux amounts depending on whether it is calculated using the flux values on its outer boundary or using field line tracing and calculating it on the photosphere. This discrepancy was first noticed when attempting to confirm that the total amount of open flux calculated on the outer boundary of WSA matched that determined using field line tracing. The two methods should give the same result, in principle, at least to within a small margin of error (e.g., round off error due to finite grid resolution, etc.), and yet they clearly do not. 

However, if one calculates the total open magnetic flux for those open cells identified for the bi-directional (i.e., orange cells in Figure 2) plus outward field line tracing method (i.e., the aqua colored cells in Figure~\ref{Fig.2} resulting from outward only tracing and that tend to reside inside coronal holes), the total open flux agrees well with that calculated on the model’s outer boundary. This result clearly reveals that the perimeter cells are the primary source producing the discrepancy. Recall that these are the cells identified as open via inward tracing only (e.g., purple cells in Figure~\ref{Fig.2}).  

The WSA model uses the adaptive step Runge-Kutta-Fehlberg (RKF) Method \citep{mathews2004} for tracing magnetic field lines, which is a relatively accurate method for tracing field lines, as it adjusts the tracing step sizes depending on how quickly the magnetic field changes. The RKF method accomplishes this by comparing the difference between the fourth and fifth order Runge-Kutta solutions and adjusting the step size to maintain the difference between the two solutions to within a specified error. In Figure~\ref{Fig.5}, we highlight two key scenarios where the field line tracing approach overestimates the amount of open magnetic flux, and therefore, explains the discrepancy between the flux calculated on the outer boundary of the model and that obtained using field line tracing.

\begin{figure}[ht!]
\includegraphics[scale=0.5] {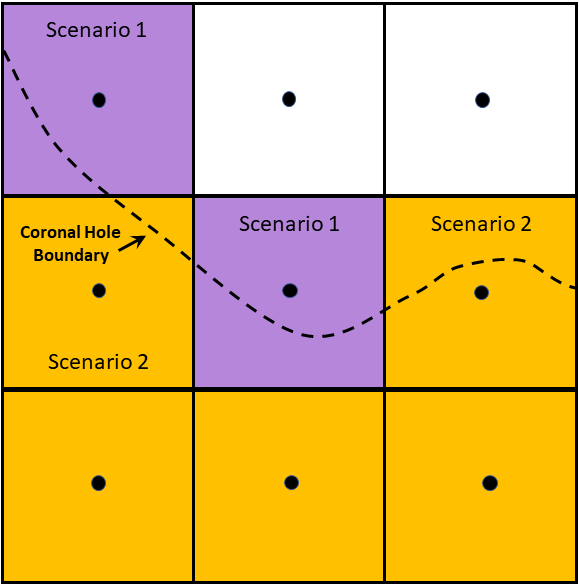}
\centering
\caption{Cartoon depiction of two key scenarios where open flux is overestimated using field line tracing. In Scenario 1, the model coronal hole boundary (dotted line) lies between the outer boundary of a bi-directional cell (i.e., orange-shaded cells where both inward and outward field line tracing identifies them as open) and the center of a cell identified as open only via inward tracing (i.e., purple perimeter cells). More than half the flux in the cell is improperly included in the open flux estimate. Scenario 2 occurs when the coronal hole boundary lies between the cell center of a bi-directional cell and the boundary of a closed cell (white cells). In this case, less than half of the flux in the cell is improperly included in the open flux estimate. 
\label{Fig.5}}
\end{figure}

In Scenario 1, the actual model boundary between the open and closed magnetic field (i.e., the coronal hole boundary) lies outside of the last cell for which both inward and outward tracing methods agree (i.e., a bi-directional cell) and the center of the neighboring perimeter cell (i.e., the one identified as open only by inward tracing). In this case, the entire magnetic flux in the perimeter cell is included in the flux calculation, when only part (less then half) of it should be. Scenario 2 occurs in a bi-directional cell where the boundary between open and closed field lines passes through it. In this case, the boundary must run between the cell center and the edge of a neighboring closed cell, as this is the only way it can be identified as open for both outward and inward tracing. Here too, the entire cell is counted when only part (more than half) of it should have been. Other possible scenarios can arise where cells are incorrectly counted as open or closed due the accumulation of small errors in the field line tracing. For instance, as a field line is traced inwardly, the accumulation of small errors can potentially lead to it landing slightly outside an open cell and just inside a neighboring closed cell causing it to be improperly counted as open. This same problem can occur for outward field line tracing where the field line, due to the accumulation of small errors, does not reach 2.49$R_{s}$ (or the source surface if just using a PFSS model) resulting in it being counted as closed. The reverse case can obviously occur too. While cases such as these likely happen, they are almost certainly rare due to the robust field line tracing method employed. In fact, Figure~\ref{Fig.2} reveals that the combined bi-directional plus outward field line tracing method provides extremely good agreement with that obtained by calculating the open flux on the outer boundary. The source of the missing flux clearly comes almost solely from the perimeter cells (i.e., the inward only tracings). This follows because if you add the flux from inward only tracings (i.e., the perimeter cells) and the outer boundary open flux seen in Figure~\ref{Fig.2}, you get approximately the in situ derived flux. 

Upon reflection, it follows that perimeter cells can provide a lot of magnetic flux if they reside near active regions. The grid resolution used for the model in this study is $2\degree$, which corresponds to about $2.4 \times 10^{4}$km at the equator. Given that active regions range in size from $10^{4}$\,--\,$10^{5}$ km \citep{tang1984}, with most regions being less than $\sim6 \times 10^{4}$km, a sizeable fraction of an active region’s magnetic flux can easily reside in a perimeter cell if it lies near an active region. Near solar maximum, the polar fields are weak and coronal holes are more common in the equatorial regions where active regions are abundant. During such times, several of the perimeter cells readily lie near active regions (see Figure~\ref{Fig.2}b) with strong magnetic fields, so these cells will have a significant amount of flux in them. The flux contributions from perimeter cells in weak field regions are generally negligible. In fact, this explains why the open flux calculated on the outer boundary of the model are in the best agreement with the in situ derived estimates during solar minimum. \citet{riley2007} found a similar result when the Wilcox Solar Observatory magnetic field data were adjusted by multiplying the values by fixed correct factor, as opposed to the saturation factor used at Mount Wilson Solar Observatory. This follows because there are few active regions along with few coronal holes in the mid-latitude regions during solar minimum. As can be seen in Figure~\ref{Fig.4}f, the polar field strengths during the 1996 minimum were about ~30\% stronger ($\sim$8 G) compared to the 2008 minimum ($\sim$6 G). Strong polar fields result in fewer equatorial coronal holes. This cycle dependence explains why the outer boundary open flux estimates agree so much better with the in situ results during the first solar minimum period shown in Figure~\ref{Fig.3} that occurred in 1996 compared to the second one in late 2008. This further argues against the notion of the missing flux being located at the poles of the Sun (e.g., \citealt{linker2017}), as this would cause the relative good agreement between the model and in situ observations to disappear during minimum and not add flux during solar maximum when the poles are weak.  

Recall that for perimeter cells, the actual model coronal hole boundary lies between a perimeter cell’s center and the neighboring bi-directional cell. So, one half or more of the cell is actually closed when the whole cell is assigned as being open, and this extra flux accounts for the discrepancy between the flux calculated using the outer boundary and the all-open cells tracing approaches. As explained above, bi-directional cells can also contribute extra flux, but their contributions to over estimating flux is generally much smaller. This discrepancy will almost certainly disappear as the model resolution is increased. In a sense, it is luck that the grid resolution used in this study happened to provide approximately the right amount of extra flux to match the in situ observations. However, this “accident” suggests a deeper explanation. PF models like WSA are static and cannot address the time evolution of the field. More advance MHD models, while time-dependent, are generally run until they obtain a steady state solution. Under this approach, the MHD models also produce sharp boundaries with the field strictly being open or close and nothing in between. However, the real Sun’s magnetic field is constantly evolving, and coronal hole boundaries represent a region where the field is constantly opening and closing. This time-dependent dynamic boundary is of some finite width that likely appears bright in the EUV. It seems as though the $2\degree$ resolution grid used in this study happens, fortuitously, to be roughly on the same spatial scale as this time-dependent boundary layer, since it is capturing the elusive missing open magnetic flux relatively well over a time span of two solar cycles. \citet{riley2003} and \citet{schwadron2005} estimated this boundary distance to be between $4\degree$ and $6\degree$, respectively. Its possible that using a coarser grid will yield an even better agreement between the outer boundary and all-cells open flux estimates. In fact, this may explain why the results of \citet{wang2022} agree reasonably well with the in situ observations. In \cite{wang2022}, the authors used a combination of low-resolution Wilcox and Mount Wilson Solar Observatory magnetograms having grid resolutions of $\sim$5$\degree$ and $4\degree$, respectively. In fact, this approach may help us better estimate the width of this time-dependent boundary. This will be explored in a future paper. 

\subsection{Absolute Calibration of Magnetographs}\label{Sec_4.3:Input_Calib}

A more detailed comparison of the total unsigned open magnetic flux as determined using in situ versus the all-open cells tracing method (cf., Figure~\ref{Fig.3}) leads to key conclusions to be drawn regarding the solar magnetographs used in this study. From 1990 to 2003, the NSO photospheric magnetic field maps used to drive the WSA model are based on the Kitt Peak Vacuum Telescope magnetograph \citep{jones1992}. In August 2003, the SOLIS/VSM instrument replaced KPVT as NSO's synoptic magnetograph. As can be seen in Figure~\ref{Fig.3}, the modeled open flux from 1990\,--\,2003 tends to slightly overestimate the in situ results. This reverses when VSM came into use in 2003, where the modeled fluxes tend to be slightly smaller than the in situ values. Several instrumental changes occurred with the VSM over the years, including: a slowly degrading modulator (between 2003 and 2006) was replaced in 2006, the spectral imaging cameras were replaced near the end of 2009, and the whole instrument was relocated in 2015 from Kitt Peak to a location in Tucson, AZ. WSA model fluxes are lower than the observed in situ values after the VSM cameras change and, especially, after the instrument relocation. It is possible that the new cameras and move resulted in the instrument’s relative calibration scaling being modified. This aside, the agreement between the WSA and the in situ results over the ~27-year comparison period is remarkably good. Assuming that the source of the missing unsigned open flux is as argued above, it then follows that the NSO line-of-sight photospheric magnetic field strength estimates are relatively accurate, as least to first order. It also implies that methods for using in situ magnetic field measurements to determine the open flux from the Sun are accurate, again to first order (e.g., \citealt{badman2021, wallace2019, owens2017} and references therein), and can be construed at the “gold standard” to be compared against (e.g., when calibrating magnetographs). 

\section{SUMMMARY} \label{Sec_5:Summary}

Since the early 2000s, the total unsigned open magnetic flux (or, for the sake of brevity, open flux) as determined by coronal models has persistently disagreed with total open flux estimates based on using in situ magnetic field measurements. Factors of two or more in  difference are seen, especially during solar maximum. Independent open flux estimates using coronal hole observations (i.e., using EUV and He \rm I} 1083.0 nm observations) and the identical input photospheric magnetic field maps used to drive the coronal models show similar disagreement with the in situ estimates and generally good agreement with those generated by coronal models (i.e., potential field and MHD, see \citealt{wallace2019,linker2017}). While differences in open flux estimates occurred prior to 2000 (e.g., 1986\,--\,1989), the differences are generally not as persistent and relatively short lived. Suggested sources of the missing flux include, for example, problems with the photospheric magnetic field measurements, incorrect polar magnetic field estimates, CME magnetic fields that have yet to disconnect from the Sun and hence counted as open when they are actually closed, the time-dependent nature of the magnetic field, and methodologies for estimating total open flux using in situ observations. 

In the study presented here, the coronal portion of the WSA model was used to calculate total open flux in two different ways. The first method makes use of the traditional approach of summing the absolute value of the magnetic flux in each grid cell on the outer boundary of the model. By construction, the magnetic field on this boundary is everywhere open. The other approach is to trace the magnetic field from the centers of each of the cells on the outer boundary down (i.e., inwardly) to the photosphere and vice versa. Grid cells on the photosphere that are determined to be open based on both outward and inward magnetic field tracing are referred to as bi-directional open cells. They represent the vast majority of the cells on the photosphere that are determined to be open in the model. Cells found to be open only based on outward tracing (i.e., from the photosphere to 2.49$R_{s}$) are referred to outward open cells. These open region cells are much fewer in number and arise due to insufficient sampling when doing inward field line tracing. Photospheric field cells exclusively open based on inward tracing are referred to inward open cells. Inward open cells are also referred to as perimeter cells, as they are found almost exclusively around the perimeters of coronal holes. As illustrated in Figure~\ref{Fig.5}, perimeter cells are open cells (purple) where the photospheric magnetic field footpoint location for inward tracing of the magnetic field line falls outside the outer boundary of a bi-directional cell (orange shading) and the center of a neighboring cell that is closed based on outward tracing exclusively. This occurs when the real model coronal hole boundary passes through this region. While only a portion (less than half) of the cell is actually open, all of it is counted using the field line tracing methodology described in this paper. The model coronal boundary can also lie between the center of a bi-directional open cell and a neighboring fully closed cell (i.e., both inward and outward tracing methods indicate the cell is closed). Here again, only part of the bi-directional cell is open (more than half), but the entire cell is counted as open due to the methodology used in this paper. These occasional additions of magnetic flux normally add little to the overall open flux calculation, as they are few in number and usually have weak magnetic field values. However, when these perimeter cells lie next to an active region, they can contribute a significant amount of open magnetic flux (refer to Figure~\ref{Fig.3}). During high activity, it can be upwards of 50\% of the total amount of open flux. Intriguingly, when these contributions are included in the model’s total open flux estimate, it produces excellent agreement with the total open flux values based on using in situ magnetic field measurements. 

Evidence presented here strongly suggests that active regions residing near the perimeters of coronal hole boundaries are the source of the solar open magnetic flux problem. There are several arguments supporting the contribution of active regions to the open flux, first, modeled and in situ derived open fluxes agree best during solar minimum when there are few active regions on the Sun and significantly diverge during solar maximum when they are abundant. The agreement between model and in situ derived open fluxes during solar the 1996 solar minimum is excellent, and while not a perfect match, shows only a slight offset for the 2009 minimum. However, the small disagreement between the open fluxes estimates during this period is a hint for the source of the discrepancy. Since the poles were $\sim$25\% weaker in 2009, compared to 1996, there were more mid-latitude coronal holes during the 2009 minimum and hence there was more access to magnetic flux from active regions (even if few in number) on the perimeters of coronal holes. This also explains why the models and observations were in better agreement prior to 2000, as the polar fields were stronger back through at least the 1970s \citep{arge2002}. Strong polar fields suppress the number of and sizes of mid-latitude coronal holes. Second, if the missing flux were in the polar fields, this would neither explain why model and in situ open fluxes during solar minimum agree so well nor account for the factors of two disagreement during solar maximum when the polar coronal holes disappear. Third, there is excellent agreement between in situ and model derived open fluxes when perimeter cells, which have high fluxes near active regions, are included. This happens during the entire 27-year time interval studied, which includes the period before the problem was apparent (i.e., before 2000) as well as afterwards. The biggest disagreement appears after 2015 and this can possibly be explained by the relocation of SOLIS/VSM from Kitt Peak to Tucson. Furthermore, active regions as the source of the missing flux also helps explain why open flux estimates based on observationally-derived coronal holes agree so well with the traditional open flux methods derived from the models, as they are bright in EUV (dark in He {\bf{\rm I}} 1083.0 nm) and therefore generally fall outside observationally-derived coronal hole boundaries. Because steady state models do not account for the time dependent evolution of the magnetic field, they assign a region as either open or closed in a binary fashion and do not account for that fact that some regions (e.g., coronal boundaries and active regions) are intermittently open. Evidence suggests these intermittently open regions have a thickness between 4\,--\,$6\degree$, and are likely bright in the EUV. As the WSA model simulations in this study used $2\degree$ resolution input maps as its input, its possible that coarser resolution maps would yield even better results. In fact, this may explain why \citet{wang2022} also got reasonably good agreement between model and in situ flux estimates, as they used coarse resolution input maps. A study where the grid resolution is varied and then compared to in situ open flux estimates could help to better determine the thickness of this intermittently open region and whether it varies over the solar cycle. 

Finally, the results presented here suggest that the magnetic field values measured using the KPVT and SOLIS/VSM instruments are generally accurate, at least to first order, and that the in situ open flux estimates can be used as a first cut constraint on the absolute calibration of photospheric magnetic field measurements.

\begin{acknowledgments}
C.N.A. is supported by the NASA competed Heliophysics Internal Scientist Funding Model (ISFM). S.W. is supported by the NASA postdoctoral program. C.J.H. is partially supported by AFOSR (Air Force Office of Scientific Research) task 22RVCOR012. The views expressed are those of the authors and do not reflect the official guidance or position of the United States Government, the Department of Defense or of the United States Air Force. This work benefited from useful discussions arising from the International Space Science Institute (Bern) team on “Magnetic Open Flux And Solar Wind Structuring Of Interplanetary Space” lead by M. Temmer (University of Graz). 
\end{acknowledgments}

\bibliography{sample631}{}
\bibliographystyle{aasjournal}



\end{document}